\documentclass[conference]{IEEEtran}
\usepackage{cite,graphicx,amssymb,amsmath,color,textcomp}
\usepackage{extarrows,multirow,multicol}
\usepackage{bm}
\usepackage{subeqnarray}
\usepackage{float}
\usepackage{subfig}
\usepackage{algorithm}
\usepackage{algpseudocode}
\usepackage{placeins}

\newtheorem{theory}{Theorem}
\newlength{\figwidth}
\IEEEoverridecommandlockouts

\begin{document}
\title{ 
Eigenvalue-Based Detection in MIMO Systems for Integrated Sensing and Communication 
\vspace{-1em}
}
\author{
 \IEEEauthorblockN{Alex Obando\IEEEauthorrefmark{1}, Saman~Atapattu\IEEEauthorrefmark{1}, Prathapasinghe Dharmawansa\IEEEauthorrefmark{2}, Akram Hourani\IEEEauthorrefmark{1}, Kandeepan Sithamparanathan\IEEEauthorrefmark{1} }
 \IEEEauthorblockA{
\IEEEauthorrefmark{1}Department of Electrical and Electronic Engineering, School of Engineering, RMIT University, Melbourne, Australia\\
\IEEEauthorrefmark{2}Centre for Wireless Communications,
University of Oulu, Finland \\
\IEEEauthorblockA{Email:\IEEEauthorrefmark{1}\{alex.daniel.obando,\,saman.atapattu,\,akram.hourani,\,kandeepan.sithamparanathan\}@rmit.edu.au}
%\thanks{This work is supported in part by the Australian Research Council (ARC) through the Discovery Early Career Researcher (DECRA) Award under Grant DE160100020 and in part by the Discovery Project (DP) under Grant DP180101205.}
%\thanks{This work is supported in part by the Australian Research Council (ARC) through the Discovery Early Career Researcher (DECRA) Award under Grant DE160100020.}
% \thanks{This work is supported in part by the Australian Research Council (ARC) Discovery Project under Grant DP220103281 and Future Fellowship under Grant FT210100728.}
}

% \thanks{
% S.~Atapattu and J.~Evans are with the Department of Electrical and Electronic Engineering,
% The University of Melbourne, Victoria, Australia (e-mail: \{saman.atapattu, jse\}@unimelb.edu.au).
\vspace{-3mm}
}
\maketitle
% \begin{abstract}

% Integrated Sensing and Communications (ISAC) has emerged as a key research area in wireless networks, improving spectral efficiency in a limited spectrum while ensuring seamless coexistence. In this work, we propose a multiple-input multiple-output (MIMO) ISAC system, where a base station (BS) equipped with multiple transmit antennas performs communication tasks. The topology includes MIMO enabled sensing and communication (S\&C) receivers (SR and CR), ensuring a fully MIMO configuration. Limited work related to MIMO ISAC has focused on performance analysis using robust analytical models. To bridge this gap, we develop a rigorous analytical framework to derive closed-form expressions for the probability of detection ($P_D$) and probability of false alarm ($P_F$) based on eigenvalue based detection. We utilize Random Matrix Theory to derive both sensing performance metrics. The communication performance is evaluated in terms of average throughput, assuming a practical scenario where the transmitter lacks full channel state information (CSI). Unlike prior works, the proposed power allocation minimizes total error probability $(P_e)$ while ensuring the average communication throughput meets a predefined threshold. Simulations confirm the validity of our theoretical analysis and highlight the effectiveness of this approach.
\begin{abstract} 
This paper considers a MIMO Integrated Sensing and Communication (ISAC) system, where a base station simultaneously serves a MIMO communication user and a remote MIMO sensing receiver, without channel state information (CSI) at the transmitter. Existing MIMO ISAC literature often prioritizes communication rate or detection probability, typically under constant false-alarm rate (CFAR) assumptions, without jointly analyzing detection reliability and communication constraints. To address this gap, we adopt an {\it eigenvalue-based detector} for robust sensing and use a performance metric—{\it the total detection error}—that jointly captures false-alarm and missed-detection probabilities. We derive novel closed-form expressions for both probabilities under the eigenvalue detector, enabling rigorous sensing analysis. Using these expressions, we formulate and solve a —{\it joint power allocation and threshold optimization}— problem that minimizes total detection error while meeting a minimum communication rate requirement. Simulation results demonstrate that the proposed joint design substantially outperforms conventional CFAR-based schemes, highlighting the benefits of power- and threshold-aware optimization in MIMO ISAC systems.

\end{abstract}

%\newpage

%\newpage

\section{Introduction} \label{S1}

% Integrated Sensing and Communication (ISAC) is a key enabler for next-generation wireless networks, offering improved spectral efficiency and tighter coordination by embedding sensing and communication (S\&C) into a unified framework \cite{9737357,9393464}. ISAC waveform design typically follows three strategies: communication-centric, sensing-centric, and joint design. While the first two approaches retrofit existing systems and may incur performance trade-offs, the joint design paradigm optimizes both functionalities from the outset, enabling superior integration~\cite{9737357}. ISAC systems may employ either a co-located sensing receiver at the transmitter~\cite{li2007} or a remotely deployed receiver~\cite{cheng2023coordinated}. The latter is advantageous when the transmitter lacks sensing capability, offering reduced hardware complexity and enabling distributed sensing in scenarios such as vehicular networks (VNETs)~\cite{qu2024near}, where multiple-input multiple-output (MIMO) systems are typically employed, either physically or virtually~\cite{nguyen2025performance}. This paper focuses on such a configuration.
Integrated Sensing and Communication (ISAC) is a key enabler for next-generation wireless networks, improving spectral efficiency by unifying sensing and communication (S\&C) functions~\cite{9737357,9393464}. ISAC waveform design typically follows three strategies: communication-centric, sensing-centric, and joint design. The joint approach optimizes both functions simultaneously for better integration~\cite{9737357}. Architectures may feature either co-located or remotely deployed sensing receivers~\cite{cheng2023coordinated}; the latter reduces hardware complexity and supports distributed sensing in systems like vehicular networks (VNETs)~\cite{qu2024near}. MIMO techniques, physical or virtual, are commonly employed~\cite{nguyen2025performance}. This work adopts such a remote sensing configuration.

\subsection{Related Works on MIMO ISAC and Motivation}

Advancements in MIMO technology have established it as a core enabler for ISAC systems, offering spatial diversity for enhanced sensing accuracy and improved communication robustness \cite{perera2025opportunistic,cui2013mimo,stoica2007probing,hua2023mimo}. Building on these advantages, recent studies have  explored MIMO-based ISAC designs. A significant  work focuses on transmit beamforming to enhance sensing by directing energy toward target directions \cite{cheng2023coordinated}. 
%In multi-user and multi-target environments, mutual information-based formulations are developed to characterize and optimize sensing metrics \cite{meng2024multi}. Joint beamforming strategies that balance both sensing and communication objectives are proposed in \cite{qu2024near}, while secure ISAC beamforming approaches are studied to mitigate eavesdropping and jamming threats \cite{liu2024joint,liu2024secure}. Moreover, reconfigurable intelligent surfaces (RIS) have been explored to enhance ISAC performance in challenging propagation conditions \cite{zhu2023joint,zhao2024dual}. 
%\com{Most existing works consider abstract utility functions such as CRLB or MI-based objectives without explicitly accounting for key detection performance metrics such as the probability of detection, false alarm, or detection signal-to-noise ratio (SNR). Constraints on communication rate, communication SNR and constant false alarm rate (CFAR) sensing are often assumed implicitly, without rigorous analysis. Moreover, practical aspects such as target availability and threshold design for required detection reliability are rarely addressed, primarily due to the lack of tractable analytical frameworks for characterizing detection performance in MIMO-ISAC settings—a problem that remains largely open. This motivates our work, which aims to bridge this gap by explicitly modeling detection performance and enabling threshold design based on concrete reliability targets.}%
In the current literature, MIMO-ISAC systems have primarily been explored through two optimization perspectives: (i) maximizing or minimizing key sensing performance metrics under communication constraints \cite{cheng2023coordinated,qu2024near,meng2024multi,zhu2023joint,Xingliang2024tcom,liu2024secure}, and (ii) improving communication performance while satisfying certain sensing-related requirements \cite{nguyen2025performance,hua2023mimo,liu2024joint,zhao2024dual}. A variety of sensing metrics have been considered, including signal-to-noise ratio (SNR) \cite{qu2024near}, mutual information \cite{meng2024multi}, and probability of detection \cite{8360142}, which remains the most widely adopted metric \cite{cheng2023coordinated,liu2024secure,Xingliang2024tcom}. On the other hand, the communication rate is often used as the principal objective function, particularly in systems aiming for guaranteed throughput \cite{nguyen2025performance,hua2023mimo}.

Notably, works such as~\cite{cheng2023coordinated,Xingliang2024tcom} apply likelihood ratio tests (LRT) with constant false-alarm rate (CFAR) assumptions~\cite{zhao2024dual}, optimizing communication SINR alongside detection probability. However, these rely on fixed thresholds, limiting sensing-communication flexibility. Generalized likelihood ratio tests (GLRT)~\cite{liu2024secure} are also common in MIMO sensing but involve costly parameter estimation and likelihood maximization, hindering scalability in joint designs.
Crucially, most prior works do not provide closed-form expressions for key detection metrics, such as false-alarm and detection probabilities. This analytical gap limits the feasibility of jointly optimizing sensing thresholds and power allocation, particularly in MIMO scenarios. While  SISO-ISAC study \cite{an2023TWC} addresses this by analyzing the trade-off between communication rate and detection probability, such methodology has not been extended to the MIMO-ISAC setting.

\subsection{Contribution}

To bridge this gap, we focus on a {\it robust and low-complexity eigenvalue-based detector (EVD)}, which is particularly well-suited for MIMO configurations with unknown signal and noise statistics. To the best of our knowledge, this is the first work to incorporate EVD into a MIMO-ISAC framework. We rigorously derive {\it novel closed-form expressions for both false-alarm and detection probabilities}—contributions that not only facilitate joint design in ISAC but also benefit general MIMO sensing applications. Leveraging these results, we propose a {\it power- and threshold-aware joint optimization} framework that minimizes the total detection error under a communication rate constraint. Unlike CFAR and GLRT-based approaches, our formulation is analytically tractable and suitable for practical implementation. Simulation results confirm that the proposed approach yields significant sensing reliability gains over existing CFAR-based designs, demonstrating its potential for effective and deployable MIMO-ISAC systems.

% Existing MIMO-ISAC studies largely prioritize communication rate under sensing constraints, commonly using CFAR-based sensing. While some adopt GLRT methods, they are computationally intensive in MIMO settings due to unknown parameter estimation and complex likelihood maximization. Moreover, most prior MIMO-ISAC works lack closed-form expressions for key detection metrics, hindering practical joint optimization of power and threshold. A recent exception \cite{an2023TWC} addresses this trade-off in a SISO-ISAC context, but its approach has not been extended to the more complex MIMO-ISAC scenario. 
% To bridge this gap, we adopt a robust and low-complexity eigenvalue-based detector (EVD), well-suited for MIMO scenarios with unknown signal and noise parameters. To the best of our knowledge, this is the first work to apply EVD within a MIMO-ISAC framework. 
% We derive novel closed-form expressions for false-alarm and detection probabilities, enabling joint system design and benefiting broader MIMO sensing applications. Using these results, we develop a power- and threshold-aware optimization framework that minimizes total detection error under a minimum communication rate constraint. 
% Unlike GLRT-based or CFAR approaches, our formulation enables practical design with tractable complexity. Simulation results show that the proposed strategy significantly outperforms CFAR-based methods in sensing reliability, highlighting its effectiveness for analytically grounded and implementable MIMO-ISAC systems.

\newpage
\section{System Model}  \label{s_system}

\begin{figure}
\vspace{-10pt} % Espacio antes de la figura
  \centering  \includegraphics[width=0.45\textwidth]{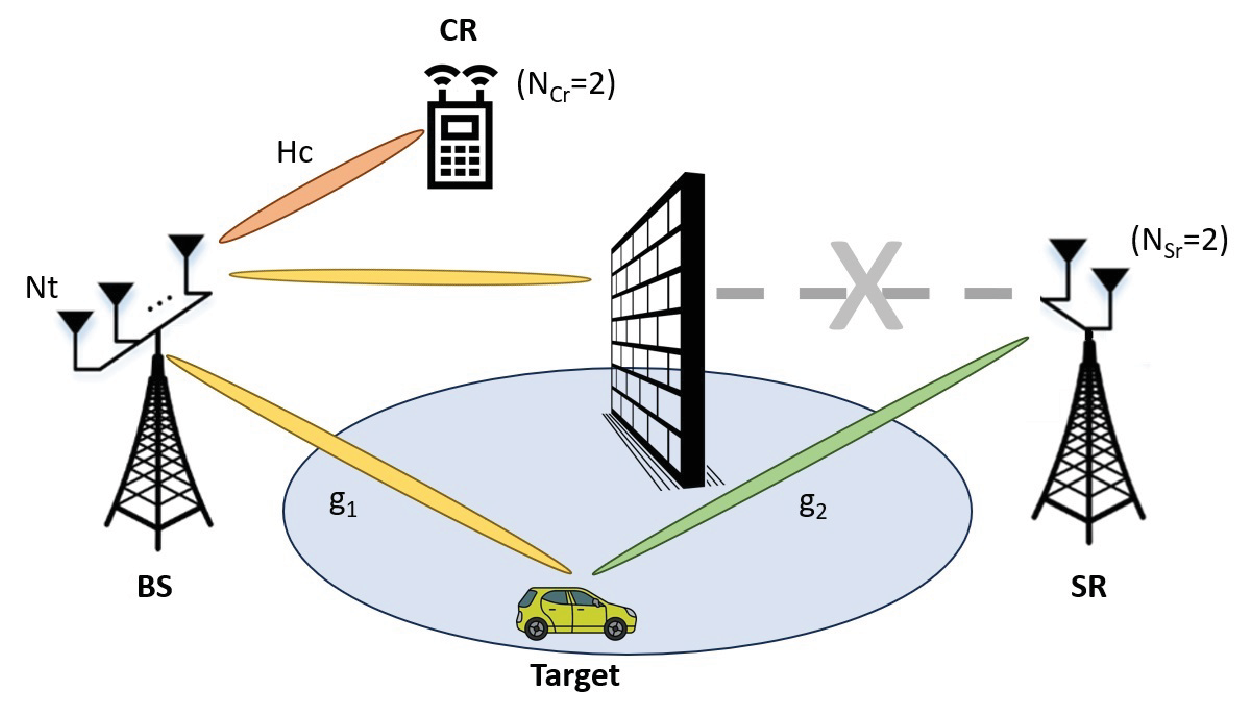}
  \caption{ISAC MIMO system model.}
  \vspace{-20pt} % Espacio después de la figura
\end{figure}\label{fig_sys_mod}
\subsection{Network Model and Signal Model}

As shown in Fig.~\ref{fig_sys_mod}, the ISAC network comprises a base station (BS) equipped with \(N_t \geq 1\) transmit antennas, which transmits a superposition of communication and sensing signals. The communication signal is intended for a communication receiver (CR) with \(N_{cr} = 2\) receive antennas, while the sensing signal enables target detection. A sensing receiver (SR)—either a standalone unit or a dual-function BS—uses \(N_{sr} = 2\) antennas to capture target echoes. This configuration constitutes a MIMO system for communication and sensing.

We assume that the sensing signal \(\mathbf{s}_s = \left[s_{s,1}, \ldots, s_{s, N_t}\right]^T \in \mathbb{C}^{N_t \times 1}\)  is known within the network, meaning that both the CR and SR have prior knowledge of it. In contrast, the communication signal \(\mathbf{s}_c = \left[s_{c,1}, \ldots, s_{c, N_t}\right]^T \in \mathbb{C}^{N_t \times 1}\) is unknown to both receivers. 
Additionally, we assume that the BS does not have channel state information (CSI), which is available only at the receivers. The total transmit power  \(P\) is shared between the sensing and communication signals, such that \(P=P_s+P_c\) both utilizing the same time-frequency resources. 
For the \(n\)th transmit antenna, the allocated power is \(P_n\), satisfying \(\sum_{n=1}^{N_t} P_n = P\). The power is divided between communication and sensing as \(\rho_{c,n}\) and \(\rho_{s,n}\), respectively, such that \(\rho_{c,n} + \rho_{s,n} = 1\). Since the BS lacks CSI, it distributes power equally among its \(N_t\) antennas, resulting in \(P_n = P/N_t\) for all \(n\). Consequently, \(\rho_{s,n} = \rho_s\) and \(\rho_{c,n} = \rho_c\) for all \(n\). Likewise \(P_c=\rho_cP\) and \(P_s=\rho_sP\). Therefore, the transmitted signal from the BS at time \(t\) can be expressed as
 \setlength{\abovedisplayskip}{2pt}
\setlength{\belowdisplayskip}{2pt}
\begin{align}\label{eq_st}
    \mathbf{s}(t) 
    & ={P_s} \mathbf{s}_s(t) + {P_c} \mathbf{s}_c(t) \in\mathbb{C}^{N_t\times 1}.
\end{align}
We assume that \(\mathbf{s}_s(t)\) and \(\mathbf{s}_c(t)\) are statistically independent and uncorrelated, satisfying \(\mathbb{E}(|\mathbf{s}_s(t)|^2) = \mathbb{E}(|\mathbf{s}_c(t)|^2) = 1\).  
\vspace{-5pt}
\subsection{Communications Signal Model}
\vspace{-4pt}
The narrowband quasi-static block fading channel from the ISAC-BS to the CR is represented by the channel matrix \(\mathbf{H}_c = [h_{ij}] \in \mathbb{C}^{2 \times N_t}\), where \(h_{ij} \sim \mathcal{CN}(0, \sigma^2)\) are independent and identically distributed (i.i.d.) complex Gaussian random variables. Then, the signal received at the CR, $\mathbf{r}_c(t)\in \mathbb{C}^{2 \times 1}$, can be expressed as
\begin{align}\label{eq_rct}
    \mathbf{r}_c(t) = \left({P/N_t}\right) \, \mathbf{H}_c (\rho_s\mathbf{s_s}(t)+\rho_c\mathbf{s_c(t)}) + \mathbf{n}_c(t)
\end{align}
where \(\mathbf{n}_c(t) \in \mathbb{C}^{2 \times 1}\) is the additive white Gaussian noise (AWGN) vector at the CR, modeled as \(\mathbf{n}_c(t) \sim \mathcal{CN}_2\left(\mathbf{0}, \sigma_c^2 \mathbf{I}_2\right)\).   

Given that the sensing signal \(\mathbf{s}_s(t)\) is known in advance to the CR, it can be mitigated using successive interference cancellation. Furthermore, as mentioned earlier, since the BS has no CSI, the transmit power is equally divided among all transmit antennas, and independent symbols are transmitted over the different antennas. Thus, the ergodic capacity for this particular MIMO system can be expressed as  
\begin{align*} 
\bar{R}(\rho_c) &= \mathbb{E}_{\mathbf{H}_c} \left[\log_2 \! \left( \det \left( \mathbf{I}_2 + \frac{\rho_c P}{N_t \sigma_c^2} \mathbf{H}_c \mathbf{H}_c^H \right) \right) \right] \text{ bps/Hz}  
\end{align*}  
where the expectation is taken over the channel $\mathbf{H}_c$. With the aid of the specific characteristics of this and the foundational contributions presented in \cite{alouini1999capacity}, an analytical closed-form expression for the ergodic capacity can be derived as  
\begin{align}\label{eq_avg_comm_rate}  
\bar{R}(\rho_c) &=  \sum_{i=0}^{2} a_i J_{N_t+i}\left( \frac{N_t \sigma_c^2}{\rho_c P} \right)  
\end{align}  
where \( a_0 = \frac{N_t}{(N_t-2)!} \), \( a_1 = \frac{-2}{(N_t-2)!} \), \( a_2 = \frac{1}{(N_t-1)!} \) and  $J_n(\mu) = (n-1)! \exp (\mu) \sum_{k=1}^n \Gamma(k-n, \mu) \mu^{n-k}$. The full derivation is omitted as it involves algebraic manipulation based on \cite{alouini1999capacity}. This expression can be easily implemented in software for real-time calculations.
\vspace{-9pt}
\subsection{Sensing Signal Model}
\vspace{-4pt}
The sensing system utilizes a MIMO configuration to effectively detect and monitor the target. Following the framework in \cite{cui2013mimo,haimovich2007}, where \( K \) samples are collected within a resource block (corresponding to the coherence time in the communication model), the received signal at the SR at the \( k \)th sampling instance, \( \mathbf{r}_s[k] \in \mathbb{C}^{2 \times 1} \), can be written as 
\setlength{\abovedisplayskip}{2pt}
\setlength{\belowdisplayskip}{2pt}
\begin{align}
\label{MIMOradar}
\mathbf{r}_s[k]=\sqrt{p}\mathbf{g}_2\mathbf{g}_1^H \mathbf{s}[k]+\mathbf{n}_s[k],\;\; k=1,2,\ldots,K,
\end{align}
where the power $p=P/(N_tN_r)=P/(2N_t)$ is normalized due to equal transmit power allocation and the 2-antenna SR.  
In this MIMO sensing setup, the transmit beam steering vector  
$\mathbf{g}_1 =\sqrt{a_t}\left[1,e^{-j\pi \sin(\theta_t)},\ldots,e^{j\pi (N_t-1)\sin(\theta_t)}\right] \in\mathbb{C}^{N_t \times 1}$ \cite{nguyen2025performance} points towards $\theta_t$ with gain $a_t$, a known quantity. The receive beam steering vector  
$\mathbf{g}_2=\sqrt{a_r}\left[1,e^{-j\pi \sin(\theta_r)}\right] \in\mathbb{C}^{2 \times 1}$ \cite{nguyen2025performance} points towards $\theta_r$ with gain $a_r$. The AWGN at the SR is modeled as $\mathbf{n}_s[k]\sim\mathcal{CN}_{2}\left(\mathbf{0},\sigma^2_s\mathbf{I}_2\right)$. 
As such, (\ref{MIMOradar}) can be written in light of (\ref{eq_st}), for $k=1,2,\ldots,K$, as
\begin{align}
\mathbf{r}_s[k]=&\sqrt{\rho_s p}\mathbf{g}_2\mathbf{g}_1^H \mathbf{s}_s[k] +\sqrt{\rho_c p}\mathbf{g}_2\mathbf{g}_1^H  \mathbf{s}_c[k]+\mathbf{n}_s[k].
\end{align}
Now keeping in mind that $\mathbf{r}_s[1],\mathbf{r}_2[2],\ldots,\mathbf{r}_s[K]$ are independent, for convenience, we may stack them to obtain the following matrix representation
\begin{align}
\mathbf{R}_s=\sqrt{\rho_s p}\mathbf{g}_2\mathbf{g}_1^H \mathbf{S}_s+\sqrt{\rho_c p}\mathbf{g}_2\mathbf{g}_1^H \mathbf{S}_c+\mathbf{N}_s
\end{align}
where $\mathbf{R}_s=[\mathbf{r}_s[1],\ldots,\mathbf{r}_s[K]]\in\mathbb{C}^{2\times K}$, $\mathbf{S}_s=[\mathbf{s}_s[1],\ldots,\mathbf{s}_s[K]]\in\mathbb{C}^{N_t\times K}$, $\mathbf{S}_c=[\mathbf{s}_c[1],\ldots,\mathbf{s}_c[K]]\in\mathbb{C}^{N_t\times K}\sim \mathcal{CN}({\bf{0}}_{1\times K}, \mu_c^2 {\bf{I}}_K)$ where this model is necessary because $\mathbf{S}_c$ is unknown to the SR, and $\mathbf{N}_s=[\mathbf{n}_s[1],\ldots,\mathbf{n}_s[K]]\in\mathbb{C}^{2\times K} \sim \mathcal{CN}({\bf{0}}_{1\times K}, \mu_s^2 {\bf{I}}_K)$. 

% In this work, we do not assume a priori target existence or enforce a constant false-alarm rate (CFAR) regime, where sensing SNR is often approximated as proportional to detection probability in  ISAC literature~\cite{9737357}. Instead, we fundamentally formulate the problem as a binary hypothesis test, derive a statistically valid detector, and rigorously analyze its false-alarm and detection probabilities. This approach ensures reliable and practical sensing performance under realistic conditions. 

\subsubsection{Hypothesis Testing Problem} 
Consequently, the sensing problem is formulated as the following binary hypothesis test:
\begin{align}
    {\bf R}_s
    =
\begin{cases} 
{\bf N}_s, & {{\mathcal {H}_{0}}},\\ 
 \sqrt{\rho_s p} \mathbf{g}_2\mathbf{g}_1^H \mathbf{S}_s+\sqrt{\rho_c p}\mathbf{g}_2\mathbf{g}_1^H \mathbf{S}_c+ {\bf N}_s, & {{\mathcal {H}_{1}}}.
\end{cases} 
\end{align}
Therefore, the distributional characteristics of the above hypotheses testing problem follows.\\
\vspace{-10pt}
\begin{align}\label{eq_Rs_H0H1}
    {\bf R}_s
    \sim
\begin{cases} 
\mathcal{CN}_{2,K}\left({\bf{0}}_{2\times K},\sigma_s^2\mathbf{I}_2\otimes \mathbf{I}_K\right), &\hspace{-2mm} {{\mathcal {H}_{0}}}\\ \mathcal{CN}_{2,K}\left(\mathbf{g}_2\boldsymbol{\alpha}_{s}^H,\left(\beta_{c}^2\mathbf{g}_2\mathbf{g}_2^H+\sigma_s^2\mathbf{I}_2\right)\otimes \mathbf{I}_K\right), &\hspace{-2mm} {{\mathcal {H}_{1}}}
\end{cases} 
\end{align}
where 
$\boldsymbol{\alpha}_{s}=\sqrt{\rho_s p}\mathbf{S}_s^H\mathbf{g}_1\in\mathbb{C}^{K \times 1}$ and 
$\beta_{c}^2=\rho_c p\mathbf{g}_1^H\mathbf{g}_1$. 
From the distributional characteristics in \eqref{eq_Rs_H0H1}, the eigenvalue detector (EVD) emerges as the preferred choice for our work. 

% key reasons~\cite{tao2019maximum}: (i) it effectively exploits the inherent low-rank signal structure created by the outer product \(\mathbf{g}_2\mathbf{g}_1^H\); (ii) it demonstrates robustness to unknown parameters such as the communication signal \(\mathbf{s}_c\); (iii) its computational complexity remains modest for the considered \(2 \times 2\) matrix case; and (iv) it maintains reliable detection performance for weak signals, outperforming conventional approaches like the energy detector (which ignores spatial correlation)~\cite{zeng2009} and the likelihood ratio test (which requires perfect knowledge of all signal parameters)~\cite{tao2019maximum}. This combination of properties makes the EVD particularly suitable for the proposed sensing scenario.

\subsubsection{Eigenvalue Detection (EVD)}
A target induces a rank-one shift in the mean and perturbs the covariance by a rank-one term. Hence, Roy's largest root, the maximum eigenvalue of the covariance matrix, is used as a test statistic due to its optimality against rank-one alternatives \cite{10619207,9103128}. Since the true covariance is unknown, the statistic is approximated using the sample covariance \cite{10437271,dissanayake2025uniform} as
\begin{align}
    \hat{\boldsymbol{\Sigma}} = \frac{1}{N} \mathbf{R}_s \mathbf{R}_s^H.
\end{align}
% , \(\hat{\boldsymbol{\Sigma}} = \frac{1}{N} \mathbf{R}_s \mathbf{R}_s^H\). 
Here, \(\hat{\boldsymbol{\Sigma}}\) follows a complex Wishart distribution because \(\mathbf{R}_s\) is matrix-variate Gaussian under both hypotheses. Specifically,\(\hat{\boldsymbol{\Sigma}}\) can be statistically charaterized as
\begin{align}
\label{WISHART}
    N\hat{\boldsymbol{\Sigma}}
    \sim
\begin{cases} 
\mathcal{CW}_{2}\left(N,\sigma_s^2\mathbf{I}_2,{\bf{0}}_{2\times 2}\right), & {{\mathcal {H}_{0}}},\\ 
\mathcal{CW}_{2}\left(N,\left(\beta^2_c\mathbf{g}_2\mathbf{g}_2^H+\sigma_s^2\mathbf{I}_2\right),\boldsymbol{\Omega}\right), & {{\mathcal {H}_{1}}}
\end{cases} 
\end{align}
where the non-centrality parameter \(\boldsymbol{\Omega}\) is defined as \(\boldsymbol{\Omega} = \|\boldsymbol{\alpha}\|^2 \left(\sigma_s^2\mathbf{I}_2 + \beta_c^2 \mathbf{g}_2 \mathbf{g}_2^H\right)^{-1} \mathbf{g}_2 \mathbf{g}_2^H\). Consequently, we use the leading eigenvalue of \(\hat{\boldsymbol{\Sigma}}\), denoted as \(\hat{\lambda}_{\max}\), as the test statistic.
distribution. 

% Under the null hypothesis, classical multivariate tools—such as matrix-variate integrals, hypergeometric functions, and zonal polynomials—can be used to characterize the distribution of $\hat{\lambda}_{\max}$. Under the alternative, however, deriving the eigenvalue density of a complex non-central Wishart matrix using complex zonal polynomials is intractable. While \cite{Ratnarajah2005} derives the joint eigenvalue density using invariant polynomials (a generalization involving two matrix arguments \cite{chikuse1986some, davis1979invariant, davis1981construction}), the resulting expressions are computationally impractical for evaluating the maximum eigenvalue distribution. Since a closed-form distribution is unavailable, we adopt a direct integration approach over the matrix-variate density \cite{davis1979invariant, mathai1997jacobians}.

\vspace{-10pt}
\section{Detection Performance}
\vspace{-4pt}
Sensing performance is evaluated via the detection and false alarm probabilities:
\begin{align}
\label{eq_pf}
P_F(\tau) & = \Pr\left(\hat{\lambda}_{\text{m}} > \tau \mid \mathcal{H}_0\right), \\
\label{eq_pd}
P_D(\tau) & = \Pr\left(\hat{\lambda}_{\text{m}} > \tau \mid \mathcal{H}_1\right),
\end{align}
where $\tau$ is a threshold typically set to satisfy $P_F(\tau) = \alpha \in (0,1)$. A target is declared present if the largest eigenvalue $\hat{\lambda}_{\text{m}}$ exceeds $\tau$. Thus, performance hinges on the cumulative distribution functions (c.d.f.s) of $\hat{\lambda}_{\text{m}}$ under both hypotheses. We first derive the c.d.f. under $\mathcal{H}_1$, as stated in the following. 
\begin{theory}\label{th_cdf_lambda_max}
The c.d.f. of the maximum eigenvalue of the scaled correlated non-central Wishart matrix $\hat{\boldsymbol{\Sigma}}$ is given, for $\beta\neq 0$ and $\boldsymbol{\Omega}\neq \bf{0}$, by
\begin{align}\label{eq_cdf_lambda_max_H1}
    F_{\hat{\lambda}_{\text{m}}|\mathcal{H}_1}(\tau)= & \frac{A(\tau) B_K (\tau)}{(K-1)}C_{K-1} \left(\frac{1}{b},\frac{a}{b^2},\tau  \right) \nonumber\\
&-\frac{A(\tau)  (K-2)!}{(\tau/\sigma_s^{2})^K}e^{-\frac{\tau}{\sigma_s^2}}C_0\left(-\frac{\beta_c^2|\bf{g_2}|^2}{b\sigma_s^2},\frac{a}{b^2},\tau\right) \nonumber \\
&+\sum_{k=0}^{K-1} \frac{A(\tau)(K-2)!e^{-\frac{\tau}{\sigma_s^2}}}{k!(\tau/\sigma_s^{2})^{K-k}} C_{k} \left(\frac{1}{b},\frac{a}{b^2},\tau  \right) 
\end{align}
where $a\!=\!\rho_s p (\mathbf{g}_1^H \mathbf{S}_s \mathbf{S}_s^H \mathbf{g}_1)  (\mathbf{g}_2^H \mathbf{g}_2)$, 
$b\!=\!\sigma_s^{2}+\rho_c p\mathbf{g}_1^H\mathbf{g}_1|\bf{g}_2|^{2}$, 
\begin{align*}
    % a & =\rho_s P (\mathbf{g}^H \mathbf{S}_s \mathbf{S}_s^H \mathbf{g})  (\mathbf{h}_s^H \mathbf{h}_s),\\
    % b & = \sigma_s^{2}+|\rho_c P\mathbf{g}^H\mathbf{g}||\bf{h_s}|^{2},\\
    A(\tau) & = \frac{e^{-\frac{a}{b}} \tau^{2K}}{\Gamma(K) \Gamma(K-1)\sigma_s^{2K} b^K} 
       \end{align*}
\begin{align*}
    B_K(\tau) & = \frac{(K-1)!}{(\tau/\sigma_s^{2})^K}   - e^{-\frac{\tau}{\sigma_s^2}} \sum_{k=0}^{K-1} \frac{(K-1)!}{k!(\tau/\sigma_s^{2})^{K-k}}
\end{align*}
\begin{align*}
    C_{k}(s,c,\tau) & = \frac{1}{(st)^{k+1}}  \sum_{\ell=0}^\infty \frac{(c/s)^\ell}{(K)_\ell \ell !} \gamma(k+\ell+1,st) 
    \end{align*}
where $\gamma(x,n)=\int_0^x z^{n-1}e^{-z}dz,\; n>0$ is the incomplete gamma function.
\end{theory}
\begin{IEEEproof}
See Appendix~\ref{apx_thm_1}.
\end{IEEEproof}
Moreover, following Khatri \cite{10619207}, we may write the c.d.f. under the null (i.e., $\beta= 0$ and $\boldsymbol{\Omega}= \bf{0}$) as
\begin{align}\label{eq_cdf_lambda_max_H0}   F_{\hat{\lambda}_{\text{m}}|\mathcal{H}_0}(\tau) = & \frac{\Gamma (K) -\gamma \left(K,\frac{K \tau}{\sigma_s^2}\right)^2}{\Gamma (K)^2} - \frac{e^{-\frac{K \tau}{\sigma_s^2}} \left(\frac{K  \tau}{\sigma_s^2}\right)^K}{\Gamma (K)} \nonumber + \\
    & \frac{e^{-\frac{K  \tau}{\sigma_s^2}} \left(\frac{K  \tau}{\sigma_s^2}\right)^{K-1}}{\Gamma (K)}\sum _{k=0}^{K-1} \frac{k!-\gamma \left(k+1,\frac{K  \tau}{\sigma_s^2}\right)}{k!}.
\end{align} 
Finally, in view of \eqref{eq_cdf_lambda_max_H1} and \eqref{eq_cdf_lambda_max_H0}, the $P_F(\tau)$ and $P_D(\tau,\rho_c)$ assume
\begin{align} 
\label{eq_pf}
P_F(\tau) & = \Pr\left(\hat{\lambda}_{\text{m}} > \tau \mid \mathcal{H}_0\right) = 1 - F_{\lambda_{\text{m}}|\mathcal{H}_0}(\tau), \\
\label{eq_pd}
P_D(\tau,\rho_c) & = \Pr\left(\hat{\lambda}_{\text{m}} > \tau \mid \mathcal{H}_1\right) = 1 - F_{\lambda_{\text{m}}|\mathcal{H}_1}(\tau).
\end{align}
Leveraging these results in \eqref{eq_pf} and \eqref{eq_pf}, we next propose a power- and threshold-aware joint optimization framework that minimizes the total detection error under a communication rate in \eqref{eq_avg_comm_rate}. 

\begin{figure*}
    \centering
    \subfloat[ROC curves when \( P = 8 \) and \( \rho_c = 0.9 \) ]{\includegraphics[width=0.33\textwidth]{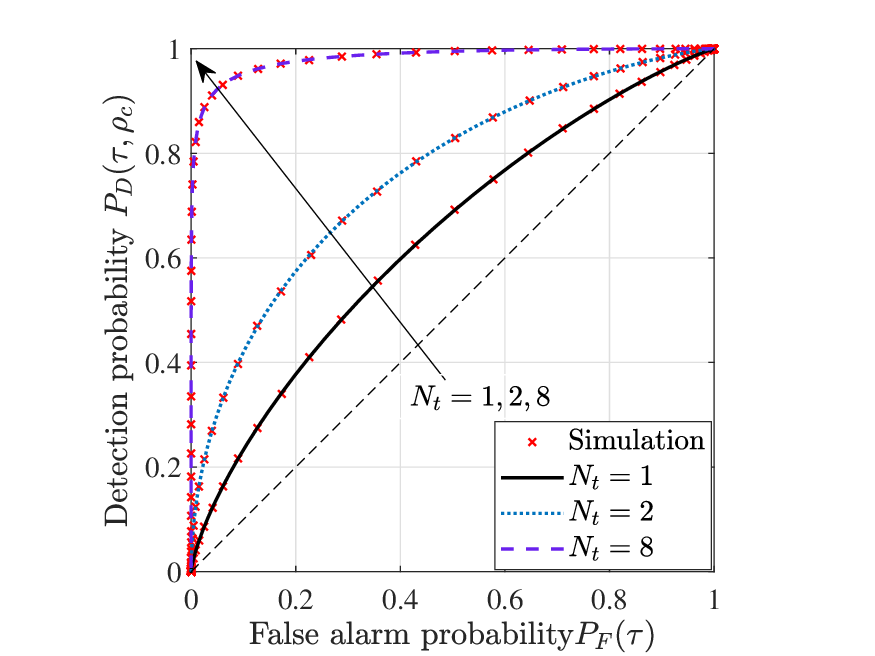}\label{fig:roc_curve}}\hfill
    \subfloat[Average communication rate vs $P_c$]{\includegraphics[width=0.33\textwidth]{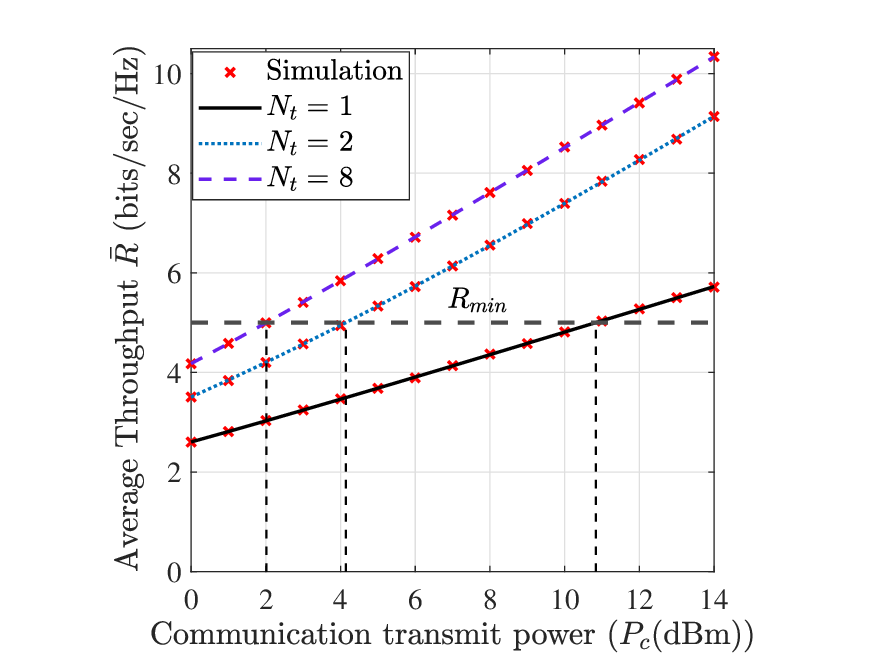}\label{fig:rate_power}}\hfill
    \subfloat[Total error vs detection threshold]{\includegraphics[width=0.33\textwidth]{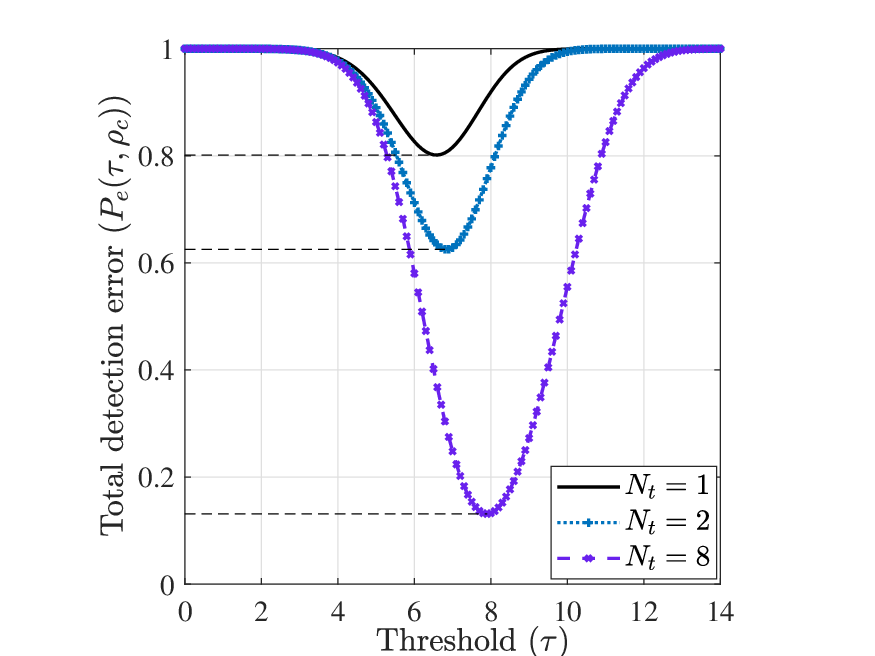}\label{fig:error_threshold}}
    \caption{Detection and communication performance for MIMO and SIMO system when \( N_t = 1, 2, 8 \).}
    \vspace{-3mm}
\end{figure*}

\begin{figure*}
    \centering
    \subfloat[False alarm probability vs $P$]{\includegraphics[width=0.33\textwidth]{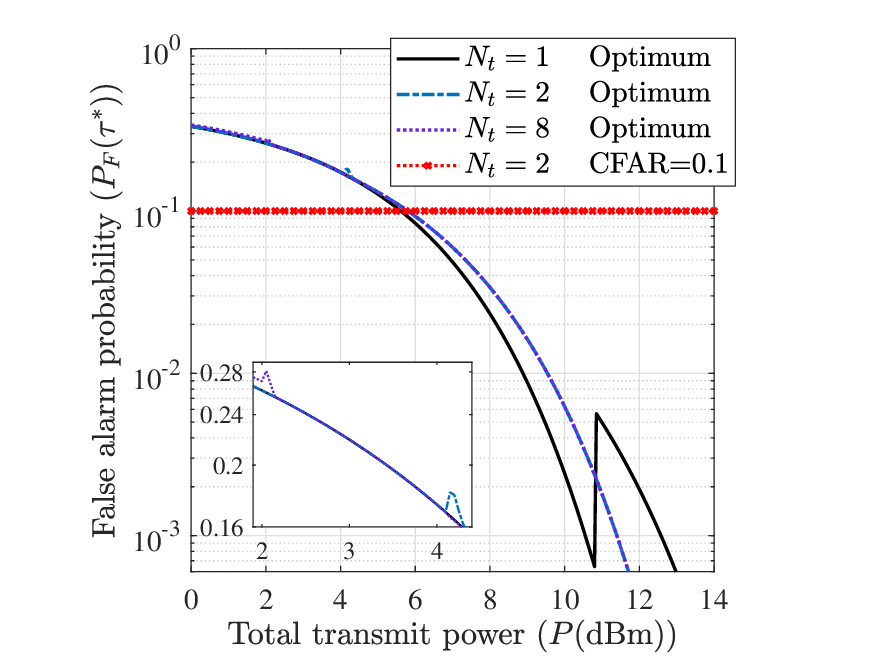}\label{fig:PFvsP}} \hfill
    \subfloat[Missed-detection probability vs $P$]{\includegraphics[width=0.33\textwidth]{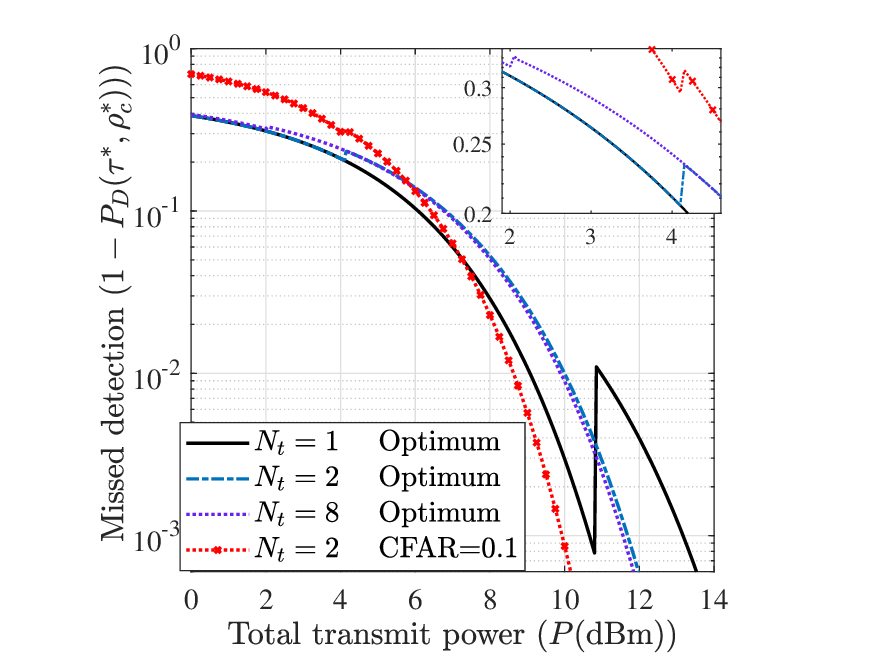}\label{fig:PDvsP}} \hfill
    \subfloat[Total error probability vs. $P$]{\includegraphics[width=0.33\textwidth]{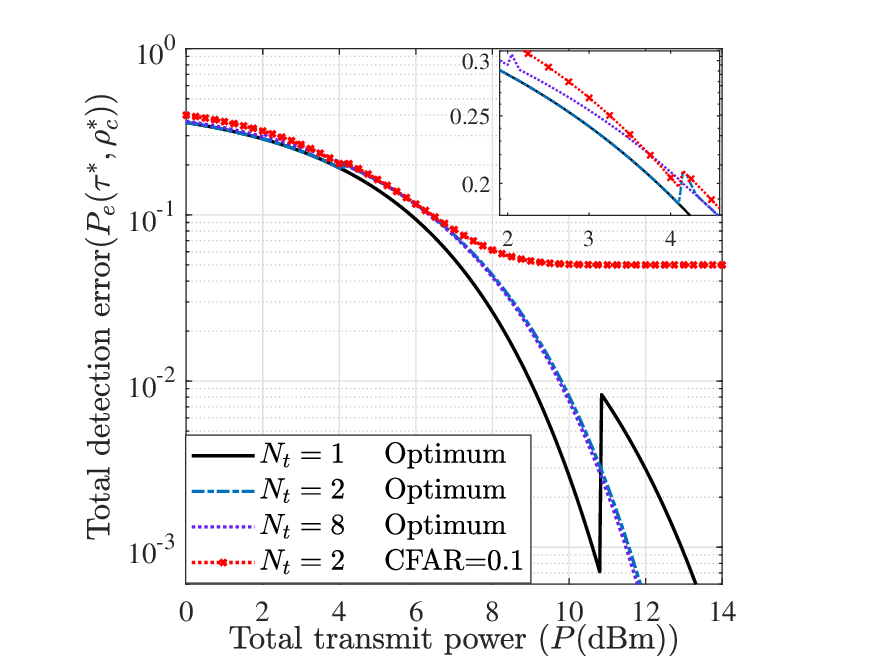}
    \label{fig:PEvsP}}
    \caption{Detection performance based on power- and threshold-aware joint design and CFAR design.}\label{fig:SenvsP}
    \vspace{-3mm}
\end{figure*}
\vspace{-3pt}
\section{Power- and Threshold-aware Joint Design}
\vspace{-3pt}
This section investigates the sensing-communication tradeoff in ISAC by jointly optimizing the power allocation $(\rho_c, \rho_s)$ across antennas and the detection threshold $\tau$ to satisfy sensing requirements. Unlike prior work~\cite{Xingliang2024tcom}, which lacks closed-form expressions, we derive explicit formulas for both throughput and detection metrics ($P_F(\tau)$, $P_D(\tau, \rho_c)$). This enables a tractable optimization problem to maximize detection performance under a rate constraint. 
\vspace{-3pt}
\subsection{Optimization Problem} 
\vspace{-3pt}
Detection performance is optimized by minimizing the total error probability,
$
P_e(\tau, \rho_c) = 1/2( P_F(\tau) + (1 - P_D(\tau, \rho_c))),
$
which balances false alarms and missed detections. This joint optimization over $\rho_c$ and $\tau$ is enabled by the closed-form expressions derived in this work, addressing a gap in existing literature and offering a novel contribution to MIMO-ISAC system design. Then, the corresponding optimization problem can be formulated as 
\begin{align}
    \min_{\tau, \rho_c} \quad & P_e(\tau, \rho_c):=0.5\left(\left( 1 - P_D(\tau, \rho_c) + P_F(\tau) \right)\right) \label{eq_opt_obj}\\
    \text{s.t.} \quad & \bar{R}(\rho_c) \geq R_{\text{min}}, \label{eq_opt_cont1}\\
    & 0 \leq \rho_c \leq 1. \label{eq_opt_cont2}
\end{align}
The objective in \eqref{eq_opt_obj} minimizes total detection error by balancing false-alarm and missed-detection probabilities. Constraint \eqref{eq_opt_cont1} ensures the communication rate exceeds a minimum threshold $R_{\text{min}}$, while \eqref{eq_opt_cont2} enforces valid power allocation. Given $\rho_c + \rho_s = 1$, we eliminate $\rho_s$, reducing the problem to a single-variable optimization over $\rho_c$. Although the total power $P$ governs allocation, it does not explicitly appear, as power is uniformly distributed across antennas as $P/N_t$.

\vspace{-3pt}
\subsection{Optimization Approach}
\vspace{-3pt}
For constraint \eqref{eq_opt_cont1}, we exploit the closed-form expression of \( \bar{R}(\rho_c) \) to derive a feasible \( \rho_c^* \):\\
\vspace{-10pt}
\begin{align}\label{eq_opt_rhoc}
    \rho_c^* =
\begin{cases} 
0, & \text{if } \bar{R}(1) < R_{\text{min}},\\ 
\bar{R}^{-1}(R_{\text{min}}), & \text{otherwise}.
\end{cases} 
\end{align}  
If full power allocation to communication ($\rho_c = 1$) fails to meet the rate constraint, all power is instead allocated to sensing. Otherwise, $\rho_c$ is set to the minimum value satisfying $\bar{R}(\rho_c) \geq R_{\text{min}}$, determined numerically by inverting \eqref{eq_avg_comm_rate}. This procedure is illustrated in Fig.~\ref{fig:rate_power} and further detailed in Section~\ref{s:num}. Having determined the feasible \( \rho_c^* \) in \eqref{eq_opt_rhoc}, we now evaluate the optimal threshold \( \tau^* \) by solving 
\begin{align}\label{eq_opt_tau}  
\min_{\tau}\, P_e(\tau, \rho_c^*) := 0.5\left(1 - P_D(\tau, \rho_c^*) + P_F(\tau)\right),  
\end{align}  
as defined in \eqref{eq_opt_obj}. In general, the total error function \( P_e(\tau, \rho_c^*) \) is convex with respect to \( \tau \), ensuring the existence of an optimal threshold \( \tau^* \) that minimizes it. This convexity can be formally verified by checking the first and second derivatives:
\begin{align}
\frac{\partial P_e(\tau, \rho_c^*)}{\partial \tau} \Big|_{\tau = \tau^*} = 0, \quad \text{and} \quad \frac{\partial^2 P_e(\tau, \rho_c^*)}{\partial \tau^2} \Big|_{\tau = \tau^*} > 0.
\end{align}
This ensures $\tau^*$ is a local minimum of $P_e(\tau, \rho_c^*)$. Although an analytical proof is possible given the closed-form expression of $P_e$, it is often cumbersome; numerical verification offers an efficient alternative. Fig.~\ref{fig:error_threshold} illustrates this, with further details in Section~\ref{s:num}. This two-step method yields feasible $\rho_c$ and $\tau$ values that solve the optimization with low complexity, enabling practical MIMO ISAC designs balancing communication and sensing.
\vspace{-14pt}
\section{Simulation Results} \label{s:num}
\vspace{-5pt}
This section presents numerical results. We consider a BS with \( N_t = 1, 2, 8 \) transmit antennas, while both the CR and SR are equipped with two antennas. The total transmit power \( P \) ranges from 0 to 14 dBm, and the noise variances are set to \( \sigma_c^2 = \sigma_s^2 = -105 \) dBm. The sensing channel gains \( a_r \) and \( a_t \) are randomly generated with path loss effects modeled using the COST-Hata model. The departure and arrival angles \( \theta_t \) and \( \theta_r \) are both set to \( \pi/6 \). Simulations are performed over \( 10^5 \) independent iterations. In this section, power and throughput values are expressed in dBm and bps/Hz, respectively.
\vspace{-8pt}
\subsection{Validation and Performance}
\vspace{-4pt}
For \( N_t = 1, 2, 8 \), with total power \( P = 8 \)  and power allocation \( \rho_c = 0.9 \) (i.e., \( P_s \approx -1.99 \) ), Fig.~\ref{fig:roc_curve} presents the ROC curves showing \( P_D(\tau, \rho_c) \) versus \( P_F(\tau) \). The analytical curves are obtained using \eqref{eq_pd} and \eqref{eq_pf}, and show a perfect match with the simulated results across all \( N_t \), thereby validating the accuracy of our closed-form expressions. As expected, the figure also confirms that detection performance improves as \( N_t \) increases. 

Fig.~\ref{fig:rate_power} shows the average communication rate \( \bar{R} \) as a function of the power allocated to the communication signal, \( P_c \), for \( N_t = 1, 2, 8 \). The horizontal line marks the rate requirement \( R_{\text{min}} = 5 \). The minimum power needed to achieve this rate is approximately \( P_c = 10.83, 4.14, 2.01 \)  for \( N_t = 1, 2, 8 \), respectively. Take \( N_t = 2 \) as an example:  
1) If the total power \( P < P_c = 4.14 \), the rate requirement cannot be satisfied, so all power is allocated to sensing.  
2) If \( P \geq P_c \), only \( P_c = 4.14 \) is allocated to communication to meet \( R_{\text{min}} \), and the remaining power is used for sensing, i.e., \( P_s = P - 4.14 \).  
This logic underpins the computation of \( \rho_c^* \) in \eqref{eq_opt_rhoc} for solving the optimization problem in \eqref{eq_opt_obj}. 
Figure~\ref{fig:error_threshold} illustrates \( P_e(\tau, \rho_c) \) as a function of \( \tau \) for different values of \( N_t \), using the same parameters as in Fig.~\ref{fig:roc_curve}. As \( N_t \) increases, the total error decreases, with minimum values \( P_{e,\text{min}} \approx 0.80, 0.62, 0.13 \) for \( N_t = 1, 2, 8 \), respectively. These minima occur at the corresponding optimal thresholds \( \tau^* \approx 6.6, 6.8, 7.9 \). This procedure is used to determine \( \tau^* \) in \eqref{eq_opt_tau} for solving the optimization in \eqref{eq_opt_obj}. 

\begin{figure}
\vspace{-10pt} 
  \centering  \includegraphics[width=0.45\textwidth]{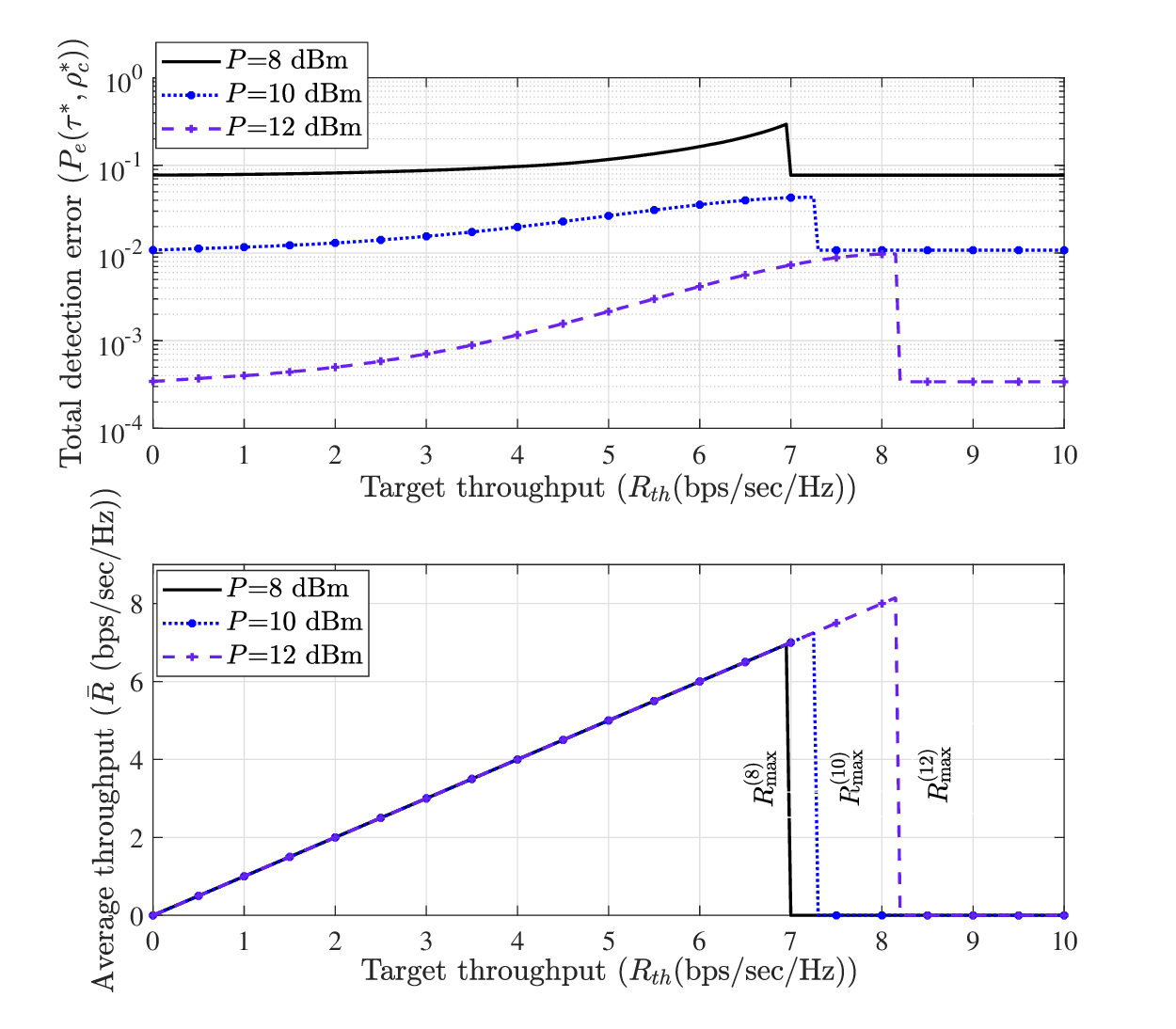}
  \caption{i) Top: \( P_e(\tau^*, \rho_c^*) \) vs target throughput \( R_{\min} \); and\\ ii) Bottom: Achieved average throughput vs \( R_{\min} \)}
  \label{fig:PevsRmin}
  \vspace{-20pt} 
\end{figure}
\vspace{-8pt}
\subsection{Optimization Approach}
\vspace{-4pt}
Fig.~\ref{fig:SenvsP} illustrates the sensing performance versus total transmit power \( P \) for \( R_{\text{min}} = 5 \) and \( N_t = 1, 2, 8 \), using the jointly optimized power allocation \( \rho_c^* \) and detection threshold \( \tau^* \) obtained from the optimization problem in \eqref{eq_opt_obj}–\eqref{eq_opt_cont2}. For comparison, CFAR-based sensing is also shown, where \( \tau \approx 5.05 \) is set to satisfy \( P_F(\tau) = 0.1 \) using \eqref{eq_pf}. Fig.~\ref{fig:PFvsP}, Fig.~\ref{fig:PDvsP}, and Fig.~\ref{fig:PEvsP} respectively show the false alarm probability \( P_F(\tau^*) \), missed detection probability \( P_{MD}(\tau^*) \), and total error probability \( P_E(\tau^*) \) as functions of \( P \). 

As seen in Fig.~\ref{fig:PEvsP}, in contrast to the CFAR design (commonly adopted in the literature), the proposed optimization consistently reduces total error as transmit power increases, confirming the {\it benefit of power- and threshold-aware joint design}. Additionally, systems with multiple transmit antennas (i.e., MIMO with \( N_t = 2, 8 \)) outperform the single-antenna case (SIMO with \( N_t = 1 \)), but only after a certain power threshold. Specifically, the SIMO system allocates all available power to sensing since it cannot satisfy the communication rate constraint \( R_{\text{min}} = 5 \); see Fig.~\ref{fig:rate_power}. Once the total power exceeds the required communication threshold (e.g., \( P \gtrsim 10.83 \) for \( N_t = 1 \)), part of the power must be reserved for communication, reducing sensing power and leading to a sharp degradation in detection performance. This explains the discontinuity observed at \( P \approx 10.83 \) for \( N_t = 1 \). Similar transitions occur at \( P \approx 4.14 \)  and \( P \approx 2.01 \) for \( N_t = 2 \) and \( N_t = 8 \), respectively, consistent with Fig.~\ref{fig:rate_power}. These results demonstrate the dual benefit of MIMO in enhancing both sensing reliability and communication capacity. 

However, the sensing performance gain between \( N_t = 2 \) and \( N_t = 8 \) is marginal. This is primarily due to equal power allocation across antennas in the absence of CSI, and the fact that the receiver employs only two antennas, limiting spatial diversity gains. With CFAR, the total error saturates at \( P_E \approx 0.06 \) as transmit power increases. For instance, at \( P = 10 \), CFAR yields \( P_E \approx 0.06 \), whereas the proposed design achieves \( P_E \approx 0.01 \) for \( N_t = 2 \), resulting in a sensing performance improvement of approximately $83$\% ($(P_{E,\text{CFAR}} - P_{E,\text{proposed}})/P_{E,\text{CFAR}}$ ).  

Similar trends are observed in false alarm and missed detection probabilities. As shown in Fig.~\ref{fig:PFvsP} and Fig.~\ref{fig:PDvsP}, the proposed design reduces both \( P_F(\tau^*) \) and \( P_{MD}(\tau^*) \) with increasing power. However, under CFAR, \( P_F \) is fixed at 0.1 by design, while \( P_{MD} \) is often lower than in our approach due to the threshold being optimized solely for false alarm control. Despite this, our objective is not to minimize \( P_{MD} \) alone, but to jointly minimize total error \( P_E = 0.5(P_F + P_{MD}) \), which provides a better trade-off between the two. Therefore, CFAR may not be suitable in settings that require adaptive, performance-driven thresholding—especially in scenarios beyond blind detection.

Fig.~\ref{fig:PevsRmin} (top) shows \( P_e(\tau^*, \rho_c^*) \) vs the target throughput \( R_{\min} \) for \( P = 8, 10, 12 \). The bottom figure plots the achieved average throughput vs \( R_{\min} \). When the power budget is sufficient, the system meets the throughput target (i.e., average throughput equals \( R_{\min} \)); otherwise, the throughput drops to zero due to infeasibility. As expected, increasing $P$ leads to lower \( P_e \), since more power can be allocated for sensing while still satisfying the communication constraint. However, as \( R_{\min} \) increases, the available power for sensing decreases, resulting in higher \( P_e(\tau^*, \rho_c^*) \). 

For instance, when \( P = 8\), the maximum achievable throughput is approximately \( R \leq R_{\max}^{(8)} \approx 6.95  \). Beyond this point (\( R_{\min} > R_{\max}^{(8)} \)), the communication requirement cannot be satisfied, and all power is allocated to sensing (i.e., \( \rho_s = 1 \)), yielding the minimum total error \( P_e^{\min} \approx 0.077 \). Similar behavior is observed for \( P = 10 \) and \( 12 \), where the corresponding throughput limits are \( R_{\max}^{(10)} \approx 7.25 \) and \( R_{\max}^{(12)} \approx 8.15 \), respectively, with associated minimum errors \( P_e^{\min} \approx 0.010, 0.00034 \). These transitions are clearly reflected in both subfigures and validate the trade-off between sensing performance and communication requirements under the joint optimization framework. 

% At low $R_{\min}$ values, the system prioritizes sensing, resulting in low detection error since only a small portion of the power is allocated to communication. As $R_{\min}$ increases, more power is allocated to communication, leading to an increase in detection error. Once the target throughput is achieved, the system allocates all the available power to sensing, and communication is effectively turned off. This behavior is reflected in the flat regions in the total detection error curves. The average achieved throughput $\bar{R}$ is shown as a function of $R_{\min}$. Initially, $\bar{R}$ increases linearly with $R_{\min}$ until the throughput requirement is met. Beyond that point, the system transitions to full sensing mode, where all power is dedicated to sensing and no power is allocated to communication. As a result, the average throughput no longer increases and drops, illustrating the boundary between joint operation and full sensing.
\vspace{-4pt}
\section{Conclusion}
\vspace{-4pt}
This paper investigated a MIMO ISAC system where a base station transmits a superimposed signal comprising a communication signal for a MIMO user and a sensing signal for a remote sensing receiver, without CSI at the transmitter. As a first in MIMO ISAC, we adopted an eigenvalue-based detector, known for its robustness in MIMO setups. Moving beyond conventional CFAR-based or $P_D$-focused designs, we introduced {\it total detection error—capturing both false-alarm and missed-detection probabilities}—as the sensing performance metric. We derived novel, closed-form expressions for these probabilities under eigenvalue-based detection, which are broadly applicable to MIMO sensing scenarios. Using these expressions and the average communication rate, we formulated a joint optimization problem to minimize total detection error subject to a rate constraint. The solution yields optimal power allocation and sensing threshold. Simulations show that our joint power-and-threshold design significantly outperforms CFAR, demonstrating effective sensing-communication balance in practical MIMO ISAC systems.

\vspace{-4pt}
\appendices
\section{Proof of Theorem~\ref{th_cdf_lambda_max}}\label{apx_thm_1} 
\vspace{-3pt}
The probability density function (PDF) of the non-central Wishart matrix \( W \) under hypothesis \( \mathcal{H}_1 \), incorporating the defined variables, is given by:
\begin{equation}
\begin{aligned}
    f(W) = &\frac{e^{-\frac{a}{b}}}{\Gamma_2(K)\sigma^{2K} b^{K}} |W|^{K-2} e^{-\frac{\text{tr}}{\sigma^2} \left[ \mathbf{I} - \frac{|\beta_c|^{2}|\mathbf{g_2}|^{2}}{\sigma_s^{2} + |\beta_c|^{2} |\mathbf{g_2}|^{2}} \mathbf{u} \mathbf{u}^{H} \right] W } \\
    &{}_0F_1\left(K, \frac{a}{b^2} \mathbf{u} \mathbf{u}^H W\right)
\end{aligned}
\end{equation}
After performing the integration, the CDF is given by:
\begin{equation}
\begin{aligned}
\label{CDF}
    F_{\lambda_m}(\tau) &= \frac{A(\tau) \pi}{(k-1)} \int_0^1 x^{K-1} e^{-\frac{\tau x}{b}} {}_0F_1\left(K, \frac{a\tau}{b^2} x\right) dx \, B_K(\tau, 1) \\
    &\quad - \frac{A(\tau)\pi}{(K-1)} e^{-\frac{\tau}{\sigma_s^2}} \frac{(K-1)!}{\left(\frac{\tau}{\sigma_s^2}\right)^K} \int_0^1 e^{\tau x \frac{|\beta_c|^2 |\mathbf{g_2}|^2}{b \sigma_s^2}} \,\\ 
    &{}_0F_1\left(K, \frac{a \tau}{b^2} x\right) dx + \frac{A(\tau) \pi}{(K-1)} (K-1)! e^{-\frac{\tau}{\sigma_s^2}} \\
    &\sum_{k=0}^{K-1} \frac{1}{k! \left(\frac{\tau}{\sigma_s^2}\right)^{K-k}} \int_0^1 x^k e^{-\frac{\tau x}{b}} \, {}_0F_1\left(K, \frac{a \tau}{b^2} x\right) dx\\
\end{aligned}
\end{equation}

The integral expression, which appears repeatedly, can be formulated as a function:
\begin{equation}
    C_k(s, c, t) = \int_0^1 e^{-stx} x^k {}_0F_1(K, c t x) \, dx
\end{equation}

Noting that ${}_0F_1(n, z) = \sum_{\ell=0}^{\infty} \frac{z^\ell}{(n)_\ell \ell!}$, we may perform term-by-term integration with some algebraic manipulation to yield
\begin{align}
C_k(s, c, t) = \frac{1}{(st)^{k+1}}  \sum_{\ell=0}^\infty \frac{(c/s)^\ell}{(K)_\ell \ell !} \gamma(k+\ell+1,st) 
\end{align}

For \( \mathcal{H}_{0} \), where \( g_2 = 0 \), \( \mathbf{S}_s = 0 \), and \( P = 0 \), the expression in (\ref{CDF}) simplifies to (\ref{eq_cdf_lambda_max_H0}).
\vspace{-3pt}

%\bibliographystyle{IEEEtran}
%\bibliography{reference,IEEEabrv}

\end{document}